\documentclass[twocolumn,showpacs,preprintnumbers,amsmath,amssymb,%
prd,a4paper,floatfix,nofootinbib,superscriptaddress]{revtex4-1}  

\usepackage{amssymb,amsfonts,amsmath,bm}
\usepackage{graphicx}

\makeatletter

\renewcommand{\@makefntext}[1]{\parindent=1em\noindent\hbox to 1.8em{\hss$^{\@thefnmark}$}#1}
\renewcommand{\@footnotemark}{\hbox{\mathsurround=0pt$^{\@thefnmark}$}}

\makeatother

\usepackage[tight]{units}

\newcommand{\eq}[1]{(\ref{#1})}
\newcommand{\fig}[1]{Fig.~{\ref{#1}}}

\newcommand{\FD}{\;.}

\newcommand{\RDH}[1]{{\mathrm{red5}(#1)}}

\newcommand{\LMH}[1]{{\mathrm{lm5}(#1)}}
\newcommand{\SU}[1]{\mathrm{SU}(#1)}
\newcommand{\U}[1]{\mathrm{U}(#1)}

\newcommand{\ubar}{\overline{u}}

\begin{document}
\date{\today}

\title{Unbreaking chiral symmetry}
\author{C. B. Lang}
\email{christian.lang@uni-graz.at}
\affiliation{Institut f\"ur Physik, FB Theoretische Physik, Universit\"at
Graz, A--8010 Graz, Austria}
\author{Mario Schr\"ock}
\email{mario.schroeck@uni-graz.at}
\affiliation{Institut f\"ur Physik, FB Theoretische Physik, Universit\"at
Graz, A--8010 Graz, Austria}

\begin{abstract}
In Quantum Chromodynamics (QCD) the eigenmodes of the Dirac operator with small absolute eigenvalues have a close relationship to the dynamical breaking of the chiral symmetry. In a simulation with two dynamical quarks, we study the behavior of meson propagators when removing increasingly more of those modes
in the valence sector, thus partially removing effects of chiral symmetry breaking. We find
that some of the symmetry aspects are restored (e.g., the masses of $\rho$ and $a_1$
approach each other) while confining properties persist.
\end{abstract}
\maketitle

\section{Motivation and Introduction}

Dynamical chiral symmetry breaking  in QCD  is
associated with the low  lying spectral modes of the Dirac operator $D$
\cite{Banks:1979yr}. They affect the path  integral weight of the gauge
configurations through the determinant of $D$. As indicated  by the
Atiyah-Singer index theorem \cite{Atiyah:1971rm}, the exact zero modes are
related  to topological excitations, the instantons. For Dirac operators
violating chiral symmetry  these are real eigenmodes. The nearby non-real modes
are also thought to be related to  composed structure of, e.g., overlapping
instantons \cite{Schafer:1996wv}.\footnote {Even when studying the low lying
modes in quenched gauge ensembles one observes  non-vanishing density and also
the Gell-Mann--Oakes--Renner relation works down to small values of the  valence
quark mass until quenched chiral logs destroy the leading chiral symmetry breaking  behavior.}

In a series of papers \cite{DeGrand:2000gq,DeGrand:2001tm,DeGrand:2003sf} it
was  emphasized that low modes saturate the pseudoscalar and axial vector
correlators at large distances and do not affect the part where high-lying
states appear.  In \cite{DeGrand:2003sf,DeGrand:2004qw} low mode saturation 
and also effects of low mode removal for mesons were studied for
quenched configurations with the overlap Dirac operator 
\cite{Neuberger:1997fp,Neuberger:1998wv}.

Subsequently low modes were utilized to improve the convergence of the
determination  of hadron propagators
\cite{DeGrand:2003sf,DeGrand:2004qw,DeGrand:2004wh,Giusti:2004yp} (see also the
recent study \cite{Bali:2009hu,Bali:2010se} comparing the efficiency when using
the low modes of the Dirac operator or the hermitian Dirac operator, where
strong dependence on the parity of the hadron states was presented). 

Associating the low mode sector with the nonperturbative chiral symmetry
breaking and the condensate \cite{Banks:1979yr}, a complementary 
question is how important it is for
confinement and mass generation of hadrons. Here we study what happens, if one
removes up to 512 low lying modes from the valence quark  sector. We compute
propagators of the  pion and other mesons and determine the effect of this 
removal on the mass spectrum. This way we want to shed light on the role of the
condensate  related to the spectral part of the Dirac operator in confinement
and chiral symmetry breaking. Our analysis is done for configurations generated
for two light, mass degenerate dynamical quark flavors. The removal of the low
lying modes is effective only in the valence quarks sector. However, as will be
seen, this already has significant impact on the meson mass spectrum.

In \cite{Glozman:2002cp,Glozman:2003bt} it has been conjectured that chiral
symmetry is ``effectively restored'' for  highly excited hadrons, in the sense
that valence quarks become less affected by the quark condensate.  This
situation is similar to ours, where we artificially suppress the condensate as
seen by the valence quarks. In the context of  effective restoration such an
approach  has been discussed already in \cite{DeGrand:2003sf,Cohen:2006bq}.

\section{Reduced Dirac operator}

Lattice Wilson Dirac operators and approximate Ginsparg-Wilson Dirac operators
are $\gamma_5$-hermitian, $\gamma_5 D \gamma_5=D^\dagger$,  but non-normal, 
thus their spectral representation has real and complex eigenvalues and the left
and right eigenvectors are bi-orthogonal, i.e.
$\langle{L_i}| R_j\rangle=\delta_{ij}$. The so-called hermitian Dirac operator
$D_5 \equiv \gamma_5\,D$ has real eigenvalues $\mu_i$  and the
eigenvectors are orthogonal.

We want to construct meson correlators from valence quark propagators which
exclude the lowest part of the Dirac spectrum. There are two alternative
definitions of reduction: based on eigenmodes of $D$ or based on eigenmodes of
the hermitian Dirac operator. We introduce the reduced quark propagator via the
spectral representation of $D_5$,
\begin{equation}\label{eq:red5}
	S_\RDH{k}=S-S_\LMH{k}\equiv S- \sum_{i\le k} \mu_i^{-1} |{v_i}\rangle\langle{v_i}|\gamma_5\FD
\end{equation}
Another alternative works with the bi-orthogonal eigensystem of $D$. The two
types of truncation are not equivalent. We first tested the convergence of the
low mode approximation and, as has been observed in \cite{Bali:2010se}, find a
clearly slower convergence rate for the standard non-hermitian as compared to the 
hermitian Dirac operator. In our study we therefore concentrate on our
results from truncating the hermitian Dirac operator.

\section{Chiral symmetry and its breaking}

The nonvanishing quark masses of the two lightest quark flavors are relatively
small in comparison to the typical QCD scale.
Neglecting the masses of the $u$ and $d$ quarks the QCD Lagrangian is invariant 
under the symmetry group
\begin{equation}
	\SU{2}_L\times\SU{2}_R\times\U{1}_V\times\U{1}_A\FD
\end{equation}
The chiral symmetry $\SU{2}_L\times\SU{2}_R$ consists of independent
transformations in the isospin space for the left- and right-handed quark fields
and can be represented equivalently  by independent isospin and axial rotations
for the combined quark fields.

The isospin axial transformation mixes states with opposite parity but the same
spin.  Depending on quantum numbers the chiral partners can have the same or
different isospin.
The non-degenerate masses of parity partners indicate the dynamical
(spontaneous) breaking of  this chiral symmetry with the order parameter
$\langle\overline \psi \psi\rangle$, the chiral  condensate. Spontaneous
breaking of the chiral symmetry leads to the appearance of the  pseudoscalar
Goldstone bosons, the pions.

The flavor singlet axial transformation symmetry $\U{1}_A$ is  broken explicitly
due to the non-invariance of the fermion integration measure, the so-called
axial anomaly. It is not a symmetry of the quantized QCD. Consequently 
no isosinglet Goldstone boson exists within the two-flavor QCD and
the $\eta$ meson(s) are heavier then the pion, attributed to the anomaly. In addition
to the anomaly also the chiral condensate breaks this symmetry.

Both symmetry breaking signals are related to low lying modes of the Dirac
operator. The axial anomaly involves the topological charge of the gauge
configuration, which is proportional to the net number of exactly chiral
(zero-)modes via the Atiyah-Singer index theorem \cite{Atiyah:1971rm}. The
chiral condensate is associated with the density of the Dirac operator's low
lying (but non-zero) modes \cite{Banks:1979yr}. The non-vanishing quark
condensate indicates breaking of both symmetries.

\section{Gauge configurations}

For our analysis we used 161 gauge field configurations \cite{Gattringer:2008vj,
Engel:2010my} of  lattice size $16^3\times 32$; with the lattice spacing
$a=\unit[0.144(1)]{fm}$ this corresponds to a spatial size of  \unit[2.3]{fm}. The simulation
includes with two degenerate flavors of light fermions and a corresponding pion
mass of $m_\pi=\unit[322(5)]{MeV}$. For the dynamical quarks of the
configurations as well as for the valence quarks   the so-called Chirally
Improved Dirac operator \cite{Gattringer:2000js, Gattringer:2000qu} has been
used. This operator is an approximate solution to the Ginsparg--Wilson equation
and therefore exhibits better chiral properties than the simpler Wilson Dirac
operator while being less expensive  by an order of magnitude -- in terms of
computation time -- in comparison to the  chirally exact overlap operator.

We calculated up to the lowest 256 eigenmodes of the Dirac operator $D$ and up to 
lowest the 512 eigenmodes of the hermitian operator $D_5$ using ARPACK which is an
implementation of the Arnoldi method to calculate part of the spectrum of
arbitrary matrices \cite{LeSoYa98}.

The quark propagator $S$ is determined by inverting the Dirac operator for a
given source. Instead of using point sources we use Jacobi smeared sources
\cite{Gusken:1989ad, Best:1997qp}  which are approximately of Gaussian shape.
Their shape was adjusted to a width of  about $\unit[0.27]{fm}$
\cite{Gattringer:2008vj}. The low mode contribution $S_\LMH{k}$ to the quark
propagator, see \eq{eq:red5}, has to be multiplied with the same sources as the full
propagator $S$ in order to achieve the correct reduced propagators $S_\RDH{k}$.

\section{Mesons}

We restrict ourselves to the study of isovectors, in particular the chiral
partners:
\begin{itemize}
\item The vector mesons $\rho$ $(J^{PC}=1^{--})$ with interpolating fields 
$\ubar(x)\gamma_i d(x)$ and $\ubar(x)\gamma_4\gamma_i d(x)$ and
$a_1$ $(J^{PC}=1^{++})$ with interpolating field $\ubar(x)\gamma_i\gamma_5 d(x)$;
in a chirally symmetric world the vector and the axial vector interpolator 
get mixed via the isospin axial
transformations.
\item The pseudoscalar $\pi$ $(J^{PC}=0^{-+})$ with interpolating fields 
$\ubar(x)\gamma_5 d(x)$  and $\ubar(x)\gamma_4\gamma_5 d(x)$. We also study the scalar
$a_0$ $(J^{PC}=0^{++})$,  $\ubar(x)d(x)$ which would get mixed with $\ubar(x)\gamma_5 d(x)$ 
via the $\U{1}_A$ transformation.
\end{itemize}
(In the interpolators $\gamma_4$ denotes the Dirac matrix in Euclidean time
direction.)

We compute from the quark propagators meson propagators, projected to vanishing 
momentum and determine the hadron masses from a range of Euclidean time values
where the correlation function exhibits exponential decay. The final errors are
statistical only and  obtained by standard jackknife elimination sampling.

\begin{figure}[t]
\includegraphics[width=0.8\columnwidth]{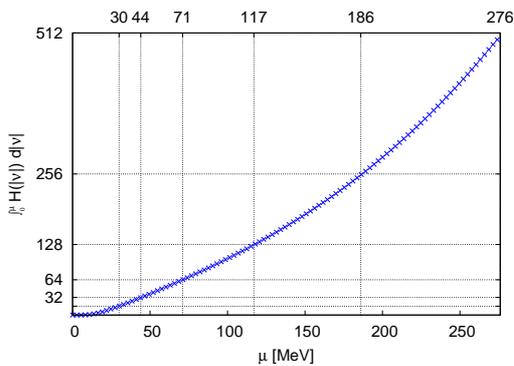}
\caption{The integrated eigenvalue density for the lowest 512 (absolute) eigenvalues of $D_5$. 
The eigenvalues are scaled according to the lattice spacing. The number on the upper axis 
indicate the values of $\mu$ where there are 16, 32, 64, 128, 256 and 512 eigenvalues below that value.
}\label{EV_integrateddensity}
\end{figure}

\begin{figure}[t]
\centering
\includegraphics[width=0.49\columnwidth]{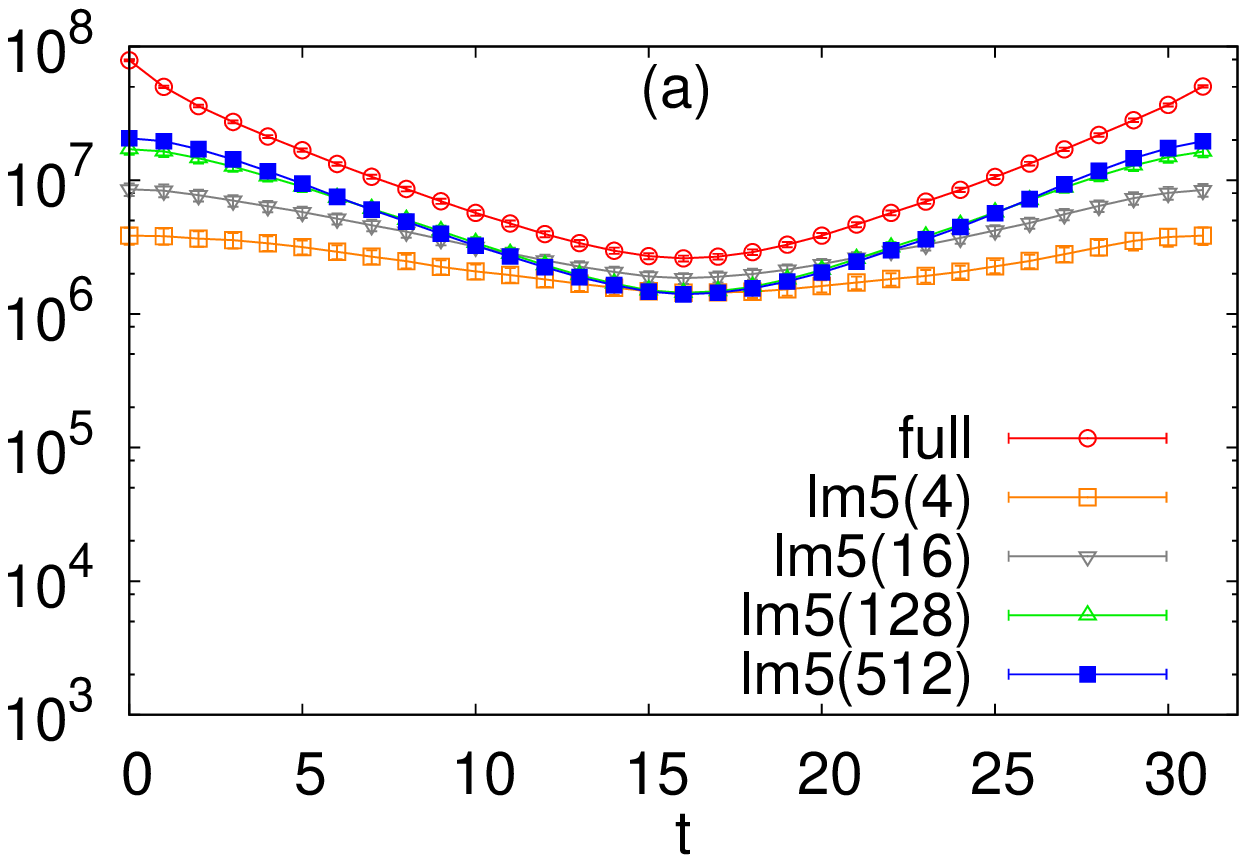}\label{0-+_LM_a}\hfill
\includegraphics[width=0.49\columnwidth]{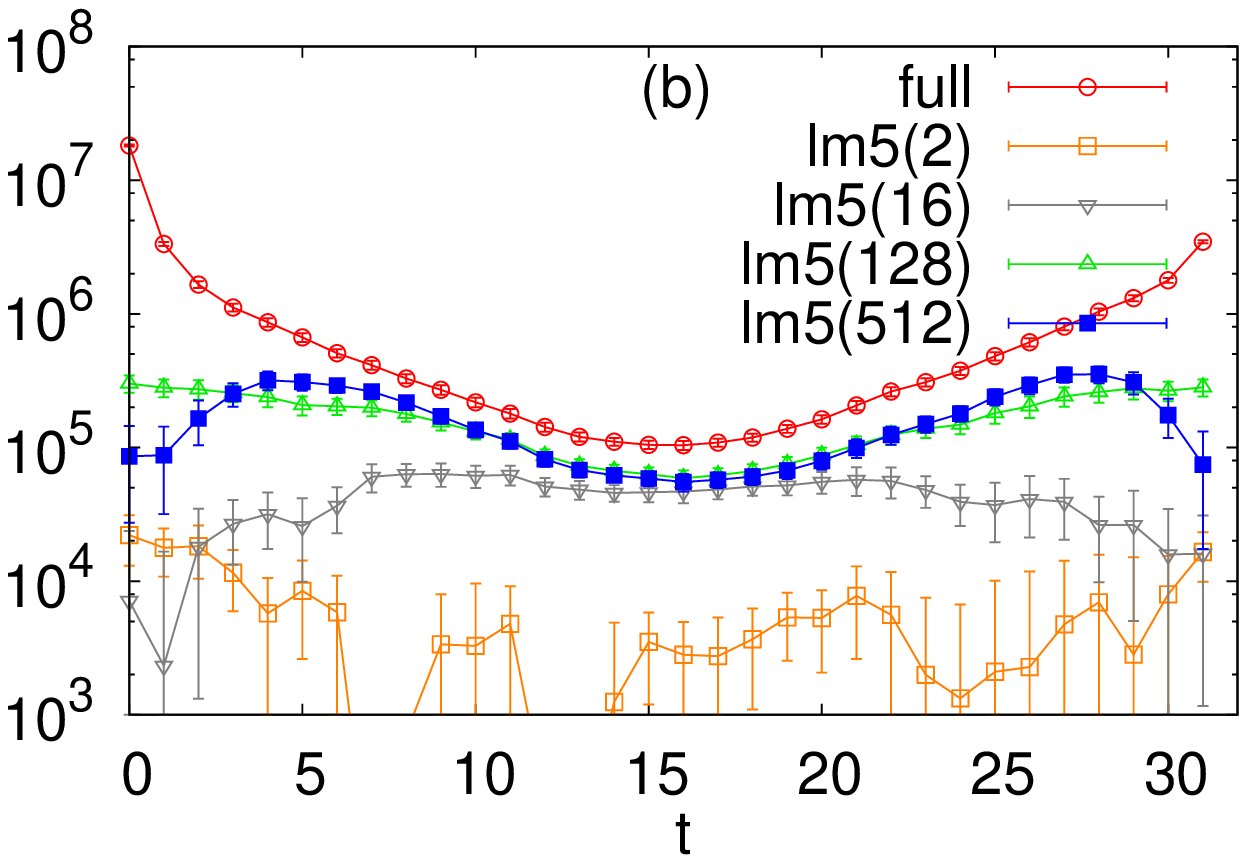}\label{0-+_LM_b} 
\caption{Low mode contribution to the correlators for the $J^{PC}=0^{-+}$ sector 
in comparison to the correlators from full propagators with interpolators
(a) $\ubar \gamma_5 d$ and (b) $\ubar \gamma_4\gamma_5 d$. The number
of included modes is shown in the legend.}\label{LM_corr}
\end{figure}

\section{Results}
\subsection{Low mode sector}

Figure  \ref{EV_integrateddensity} shows the integral over the distribution $H(|\mu|)$ 
of the (real) eigenvalues of $D_5$. 
The scale is set by the lattice spacing. There is a transition
region up to roughly twice the size of the quark mass (for this simulation the
unrenormalized mass calculated from the axial Ward identity is \unit[15]{MeV}
\cite{Engel:2010my}) corresponding to $\mathcal{O}(16)$  eigenmodes,
as also observed in, e.g., 
\cite{Luscher:2007se,Giusti:2008vb,Necco:2011vx,Splittorff:2011bj}.
As will be seen below, this is in accordance with the behavior
observed for the meson propagators.

For the overlap operator the real eigenvalues correspond to exact chiral modes,
the zero modes. This is no longer true for Wilson-type operators. There one may
associate zero modes with  real eigenvalues, although there chirality is not
unity. For the hermitian Dirac operator there  is no simple method to identify
these would-be zero modes and thus all we can say is that  the lowest 8 modes
include a significant number (if not all) of the would-be zero modes
(instantons). 

Before we construct meson correlators out of reduced quark propagators, let us
first consider meson correlators approximated by the lowest $k$ modes only, using
propagators $S_\LMH{k}$, see \eq{eq:red5}.

In \fig{LM_corr} we compare the pseudoscalar correlator using standard full
propagators to  the correlators using only the lowest modes of the hermitian
Dirac operator $D_5$. For the two  pseudoscalar operators the exponential pion
decay behavior sets it much earlier (at lower  numbers of eigenmodes) for the
interpolator $\ubar \gamma_5 d$ than for the other interpolator $ \ubar \gamma_4
\gamma_5 d$. Clearly the first one is stronger dominated by the low lying modes 
than the second. The large time region is being well described by the low modes
whereas the  short time region -- where excited states dominate -- gets much slower
saturated. Comparing  with  the result for an equivalent approximation for the
non-hermitian Dirac operator (not shown here) we find that less eigenmodes of
$D_5$ are needed to obtain a similar quality of approximation of the correlators
with full propagators. These results agree with the observations in 
\cite{DeGrand:2003sf,DeGrand:2004qw,Bali:2010se}.

\subsection{Removing the low mode sector}

\begin{figure}[t]
\centering
\includegraphics[width=0.49\columnwidth]{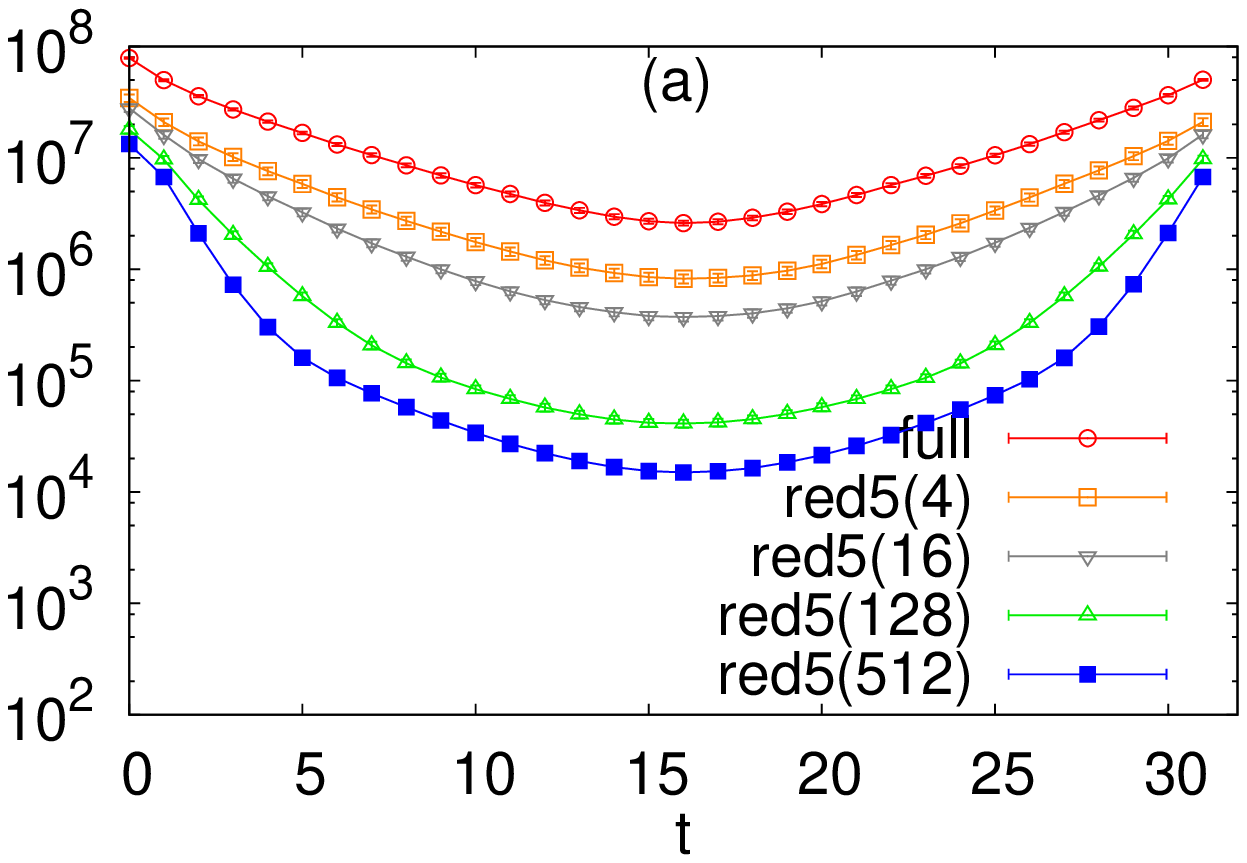}\label{0-+_RD_a}\hfill
\includegraphics[width=0.49\columnwidth]{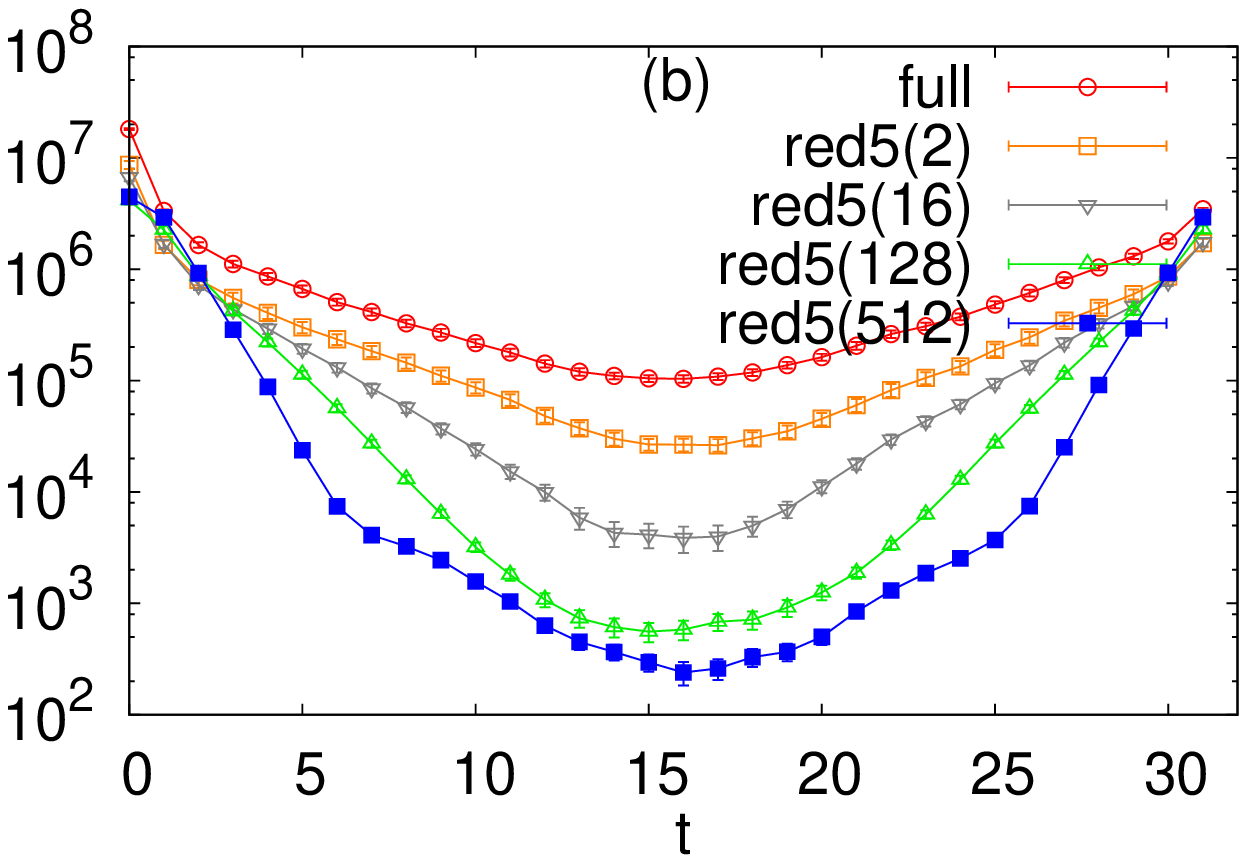}\label{0-+_RD_b} \\
\includegraphics[width=0.49\columnwidth]{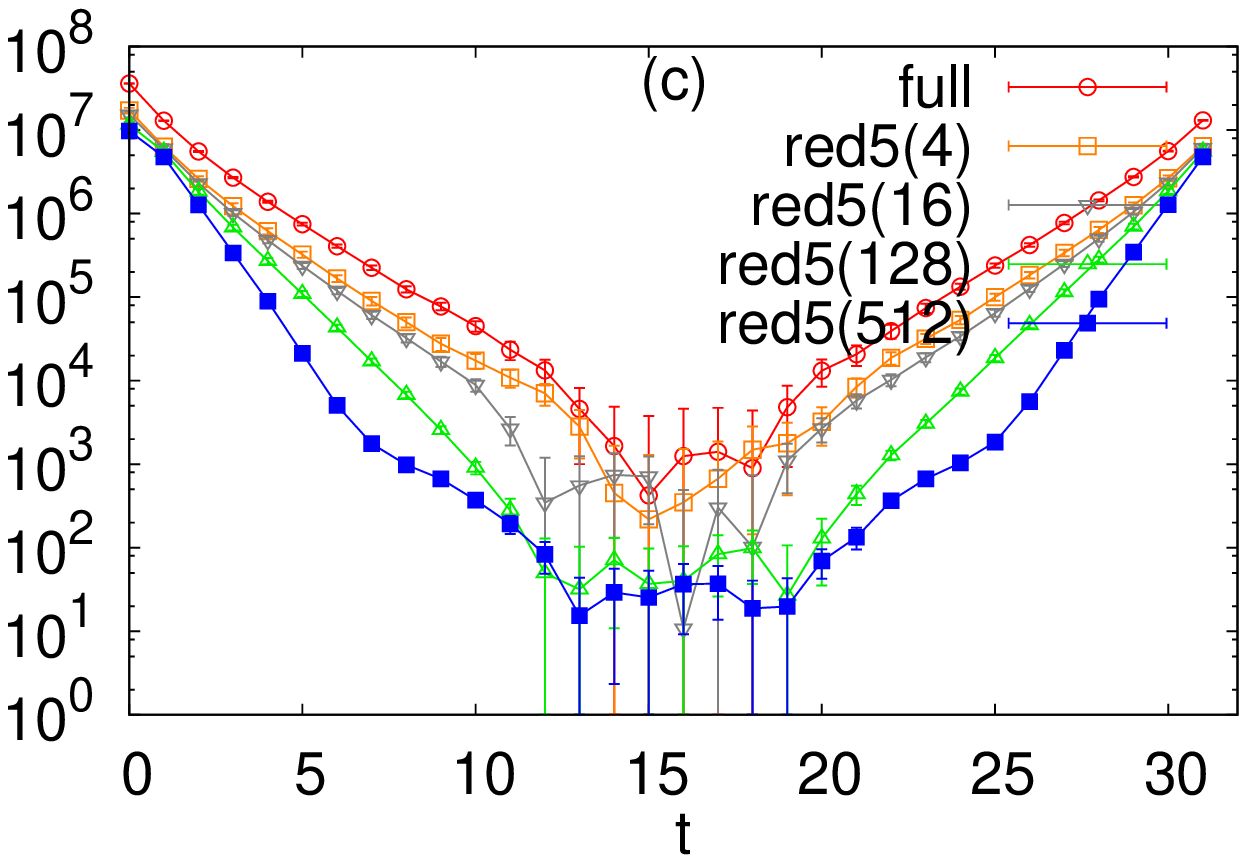}\label{1--_RD_a}\hfill
\includegraphics[width=0.49\columnwidth]{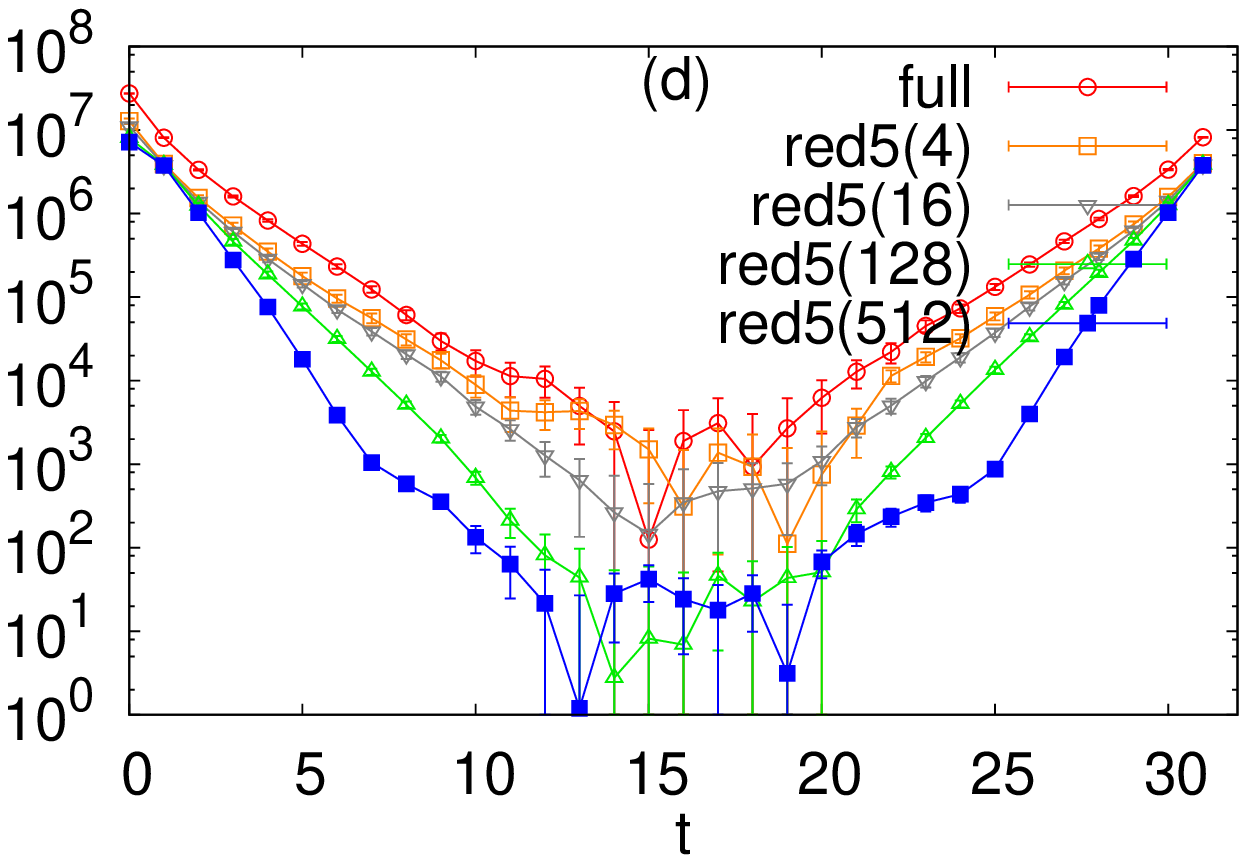}\label{1--_RD_b}\\
\includegraphics[width=0.49\columnwidth]{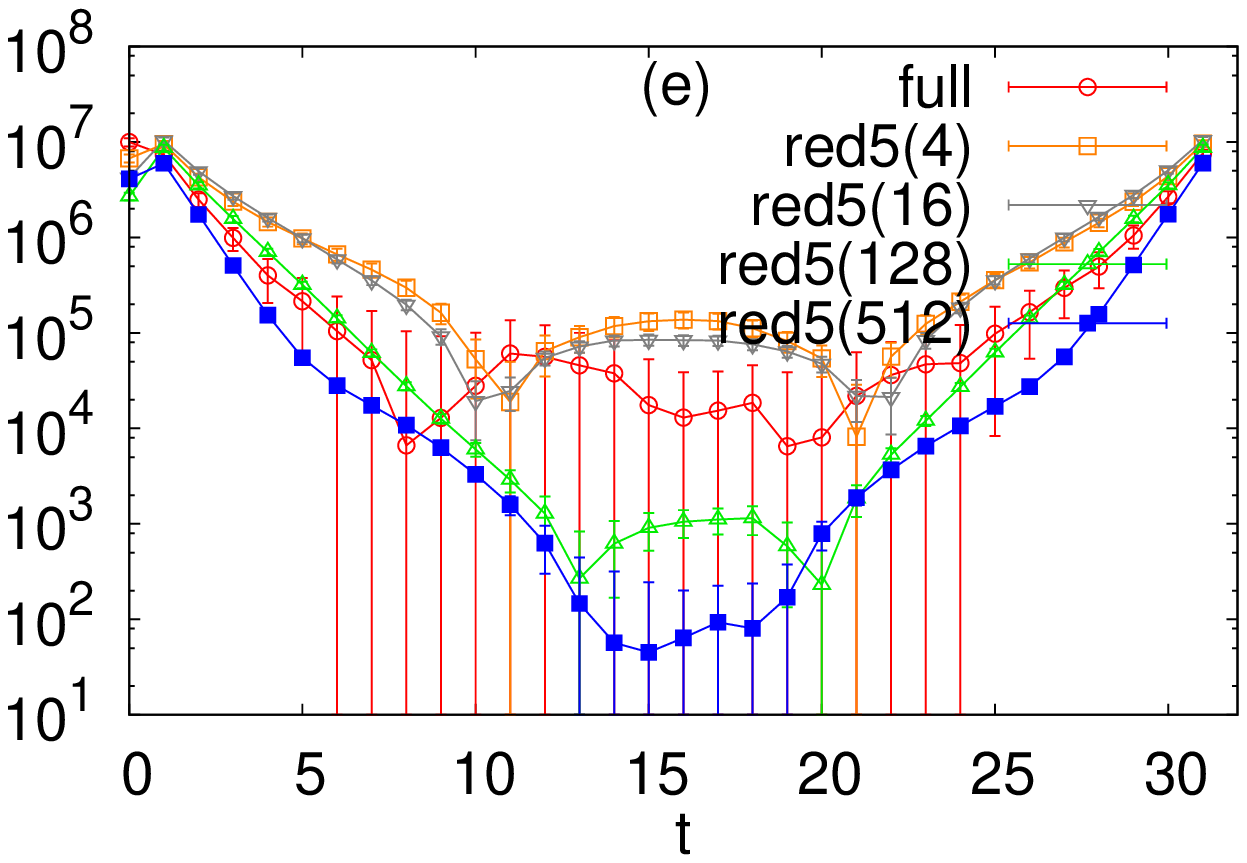}\label{0++_RD}\hfill
\includegraphics[width=0.49\columnwidth]{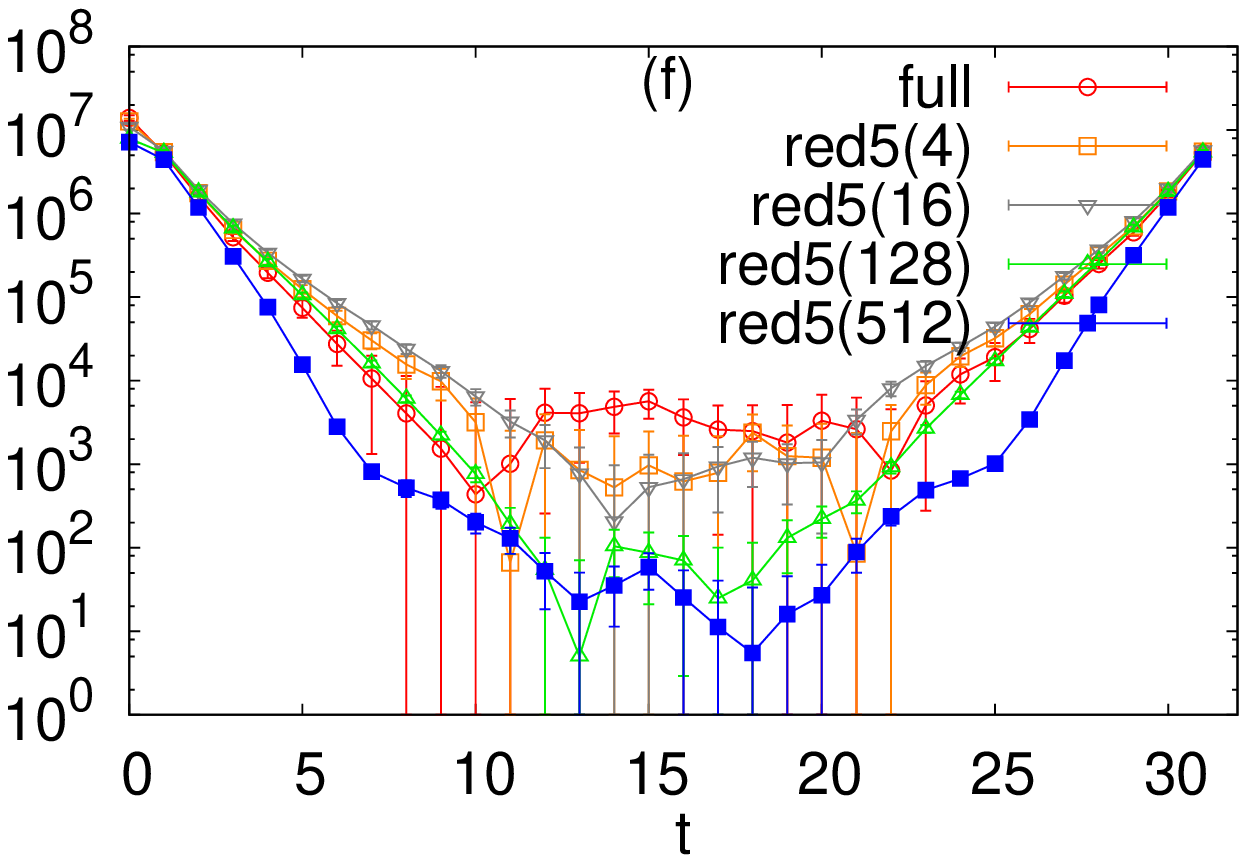}\label{1++_RD}
\caption{Correlation functions for the {\em reduced} interpolators as compared 
to the correlators from full propagators.
Top: $J^{PC}=0^{-+}$ with interpolators (a) $\ubar \gamma_5 d$, (b) $\ubar \gamma_4\gamma_5 d$.
Middle: $J^{PC}=1^{--}$ with (c) $\ubar \gamma_i d$, (d) $\ubar \gamma_4\gamma_i d$.
Bottom: Reduced (e) $J^{PC}=0^{++}$  ($\ubar d$) and
(f) $J^{PC}=1^{++}$ ($\ubar \gamma_i\gamma_5 d$).
}\label{RD_corr}
\end{figure}

\begin{figure}[th]
\includegraphics[width=1.0\columnwidth]{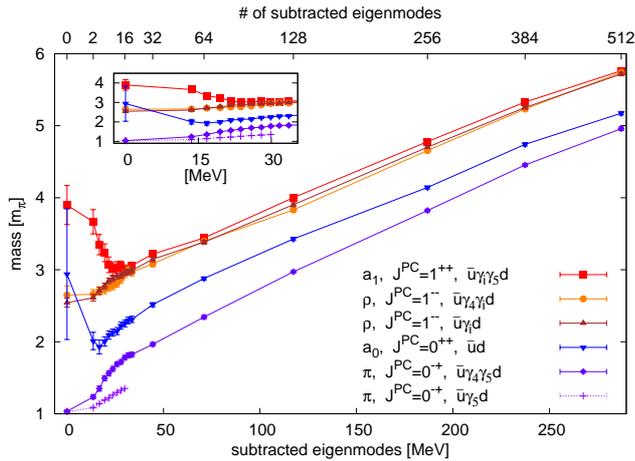}
\caption{The masses of all considered mesons as a function
of the reduced spectrum, subtracting the 0--512 lowest modes of $D_5$.
}\label{fig:masses}
\end{figure}

Figure \ref{RD_corr} shows the meson propagators for various stages of low mode
{\em removal}, always in comparison with the full propagator and \fig{fig:masses}
combines the corresponding mass fits to the regions of exponential behavior.

\begin{figure}[t]
\centering
\includegraphics[width=0.49\columnwidth]{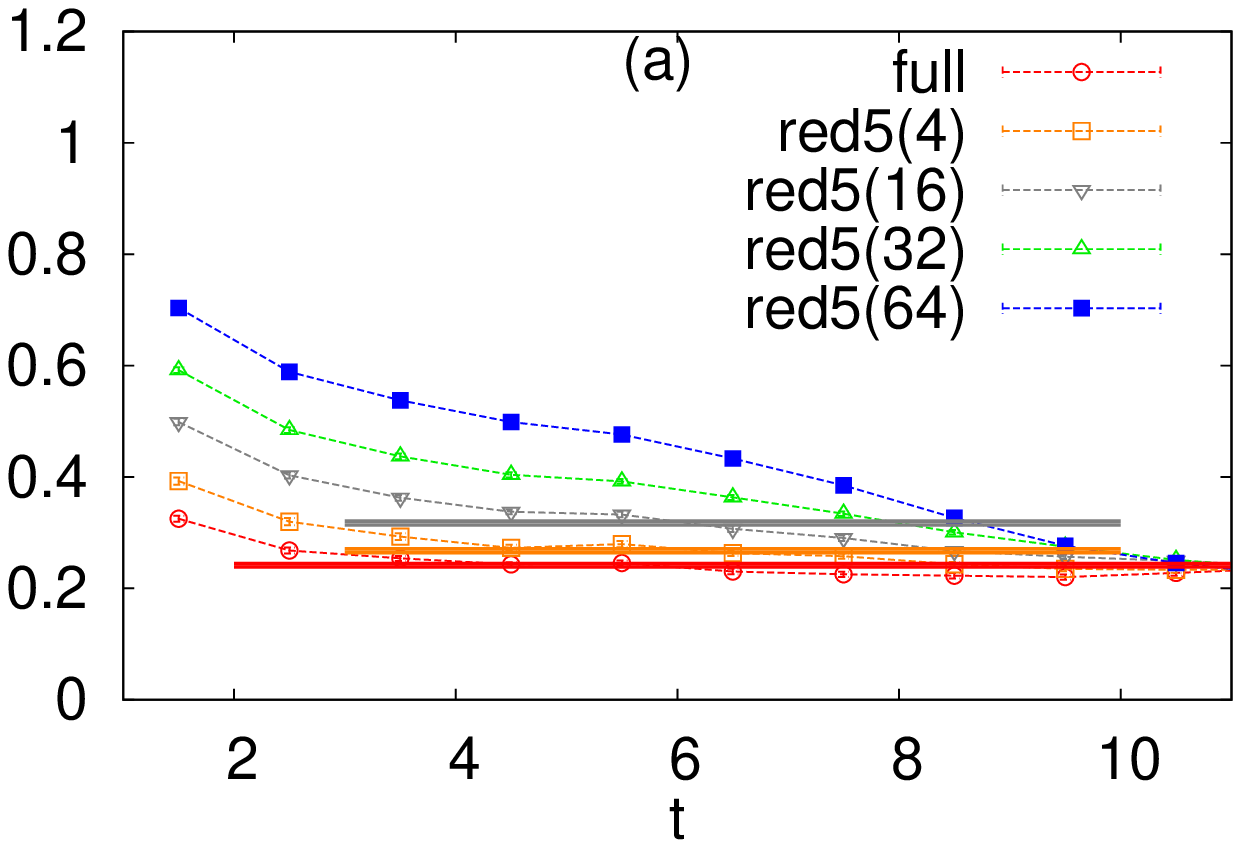}\label{0-+_RD_a_meff}\hfill
\includegraphics[width=0.49\columnwidth]{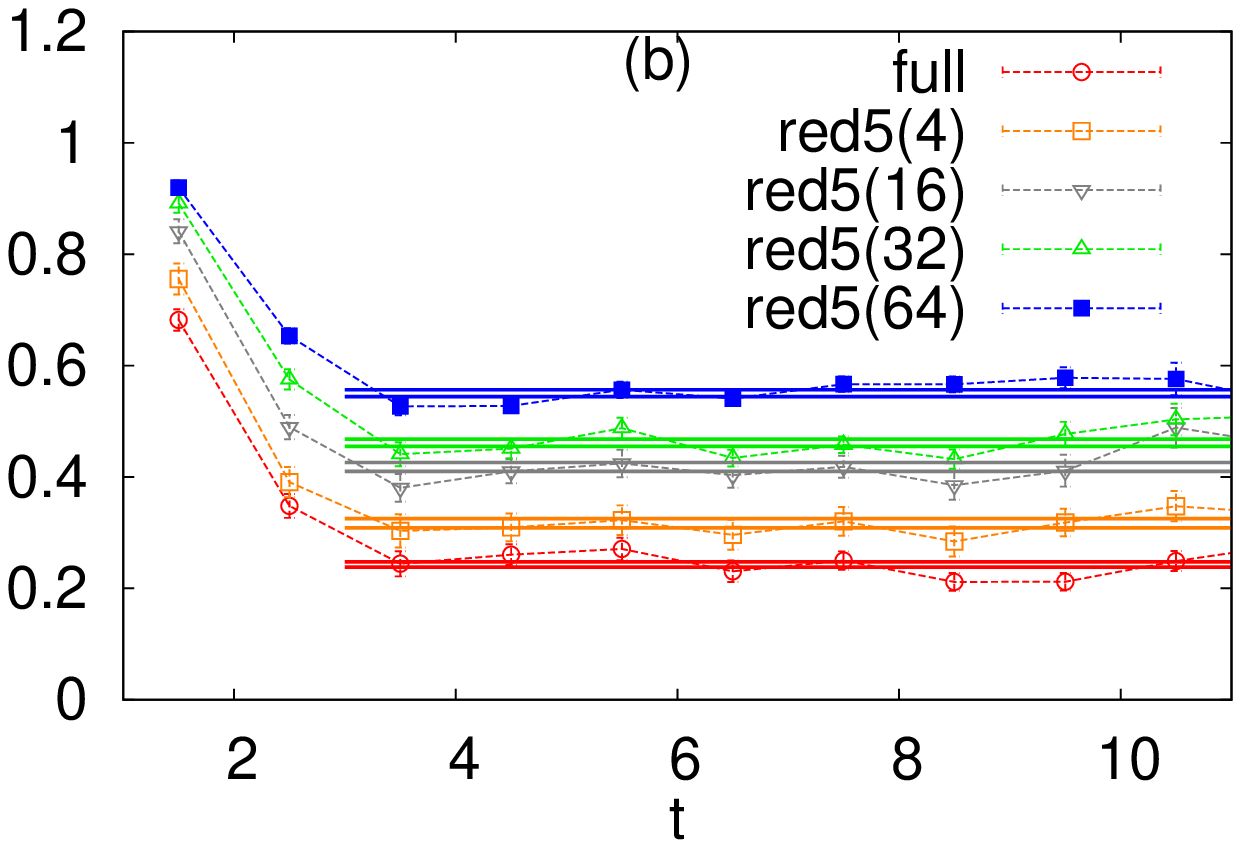}\label{0-+_RD_b_meff} \\
\includegraphics[width=0.49\columnwidth]{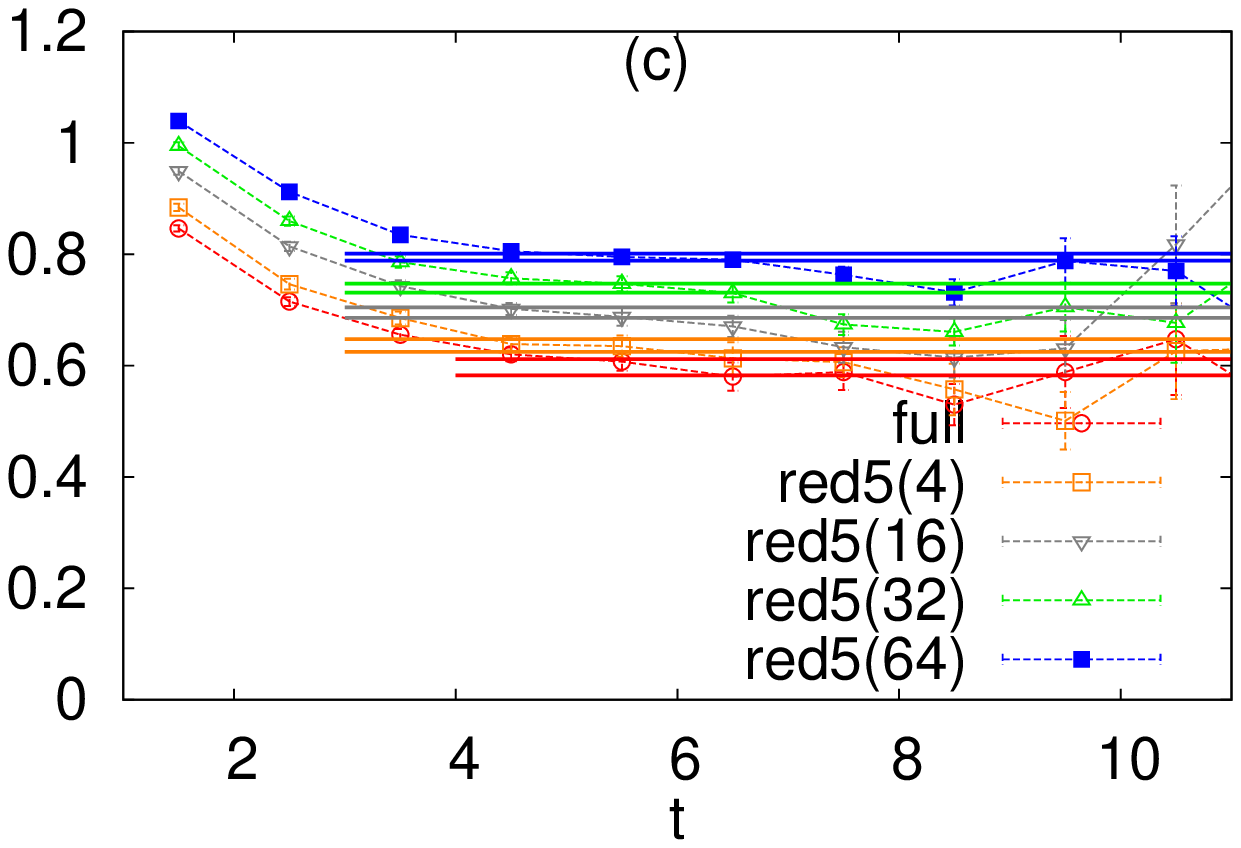}\label{1--_RD_a_meff}\hfill
\includegraphics[width=0.49\columnwidth]{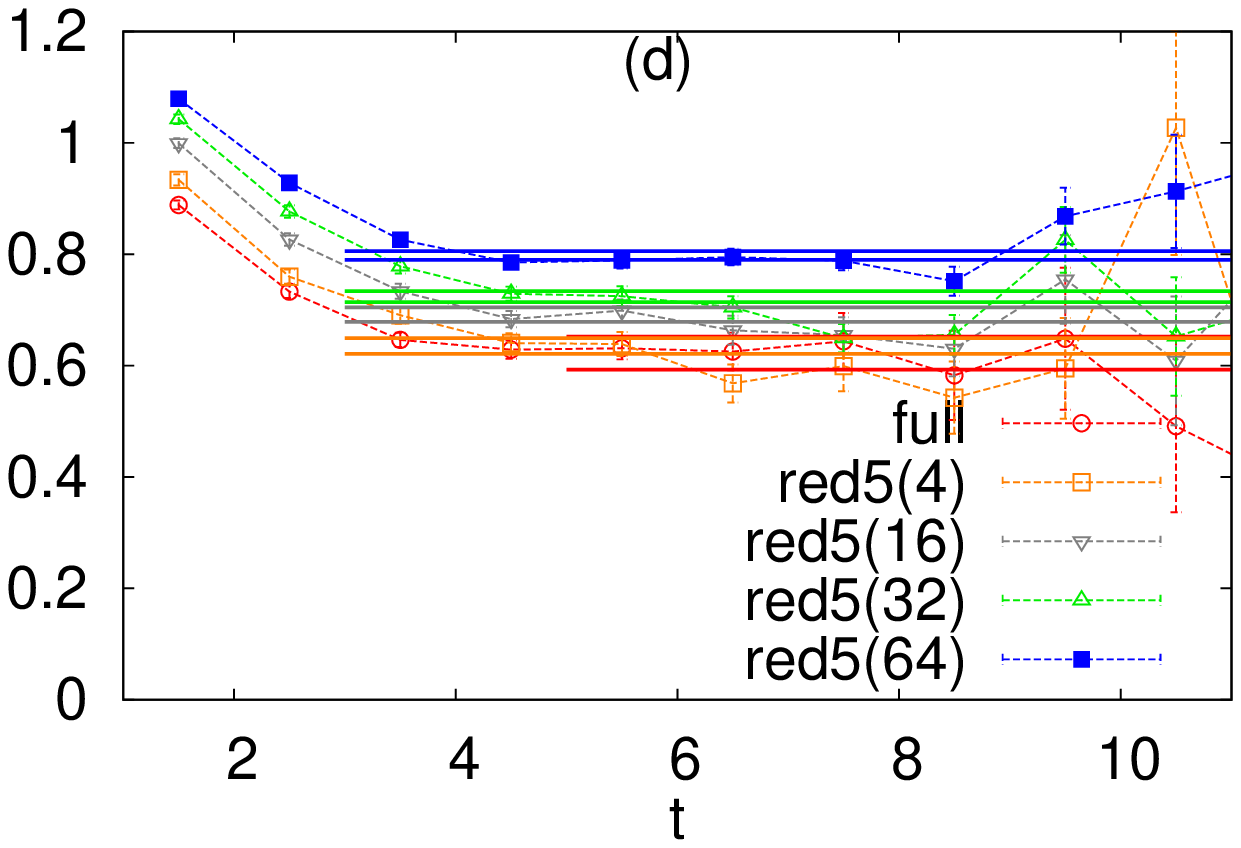}\label{1--_RD_b_meff}\\
\includegraphics[width=0.49\columnwidth]{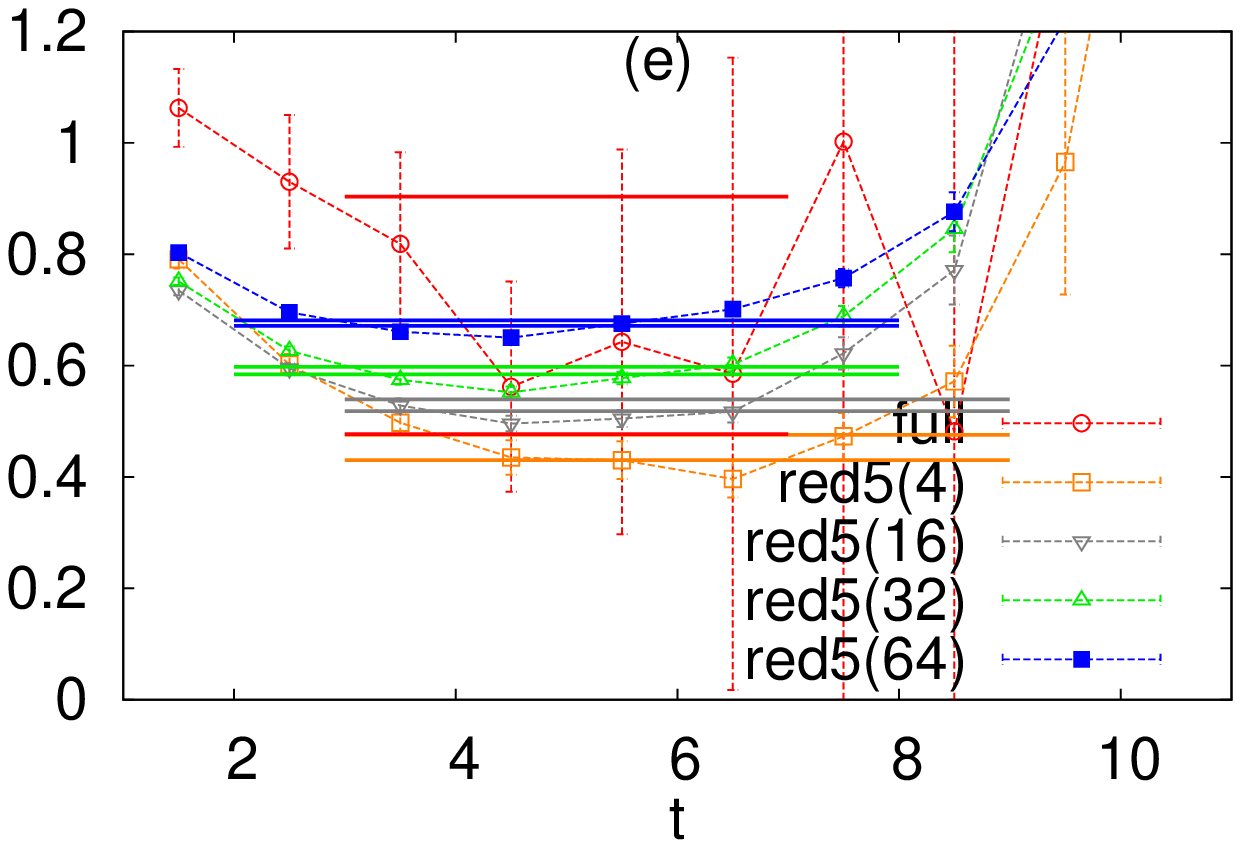}\label{0++_RD_meff}\hfill
\includegraphics[width=0.49\columnwidth]{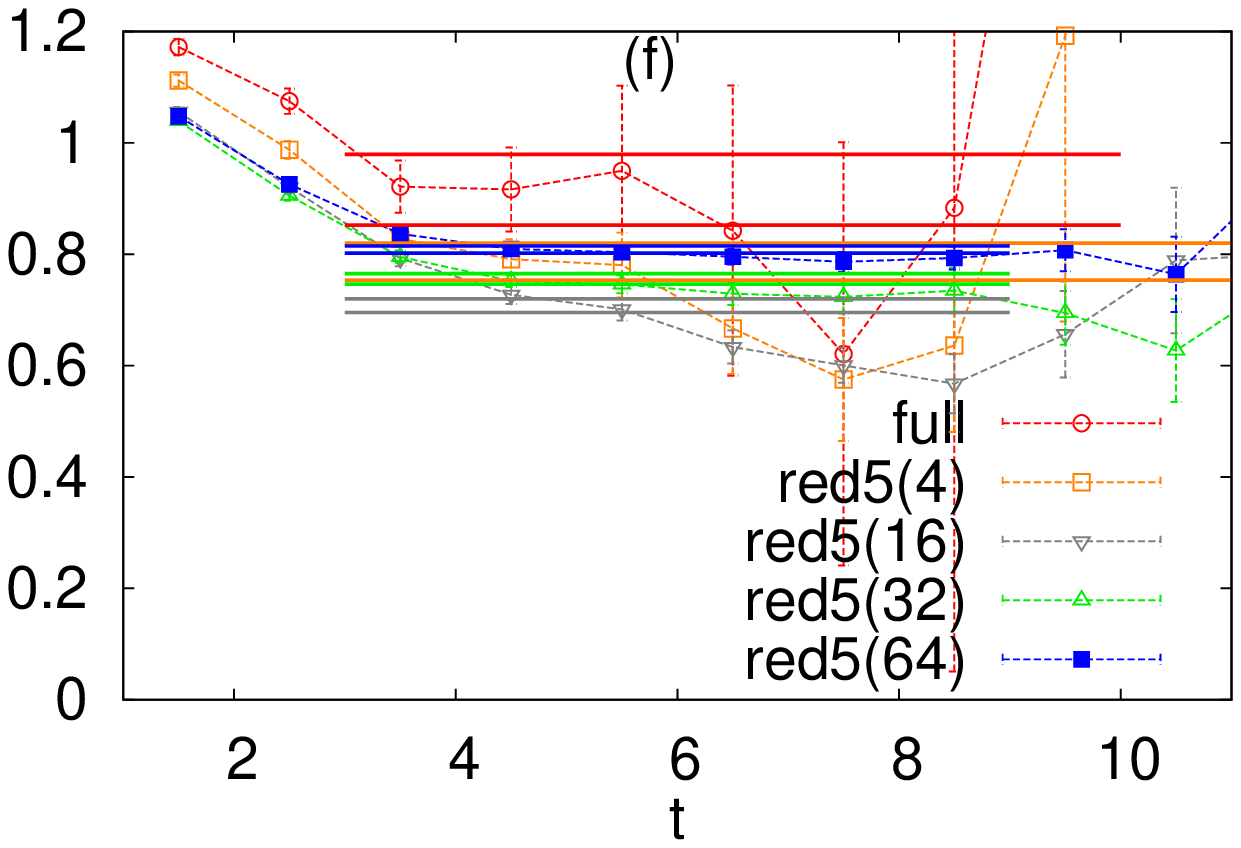}\label{1++_RD_meff}
\caption{Effective mass plots for the {\em reduced} interpolators as compared 
to the full propagators. For the notation (a-f) see \fig{RD_corr}.
}\label{RD_meff}
\end{figure}

All mass values (except for the $\rho$) exhibit a strong dependence on the
truncation of the lowest eigenmodes; from truncations levels of $\sim16$  modes
upwards (corresponding to  quark masses of approximately \unit[30]{MeV})
all mass values
then follow a roughly parallel,  rising behavior. The range of exponential
behavior of the correlators shrinks, as can be seen in the log-plots in
\fig{RD_corr}.

The effective mass plots (the local two-point approximation of the derivative of the
logarithm of the correlators) in \fig{RD_meff} indicate the regions, where an exponential
fit to the correlators has been done. We find that the fluctuation typically decreases
with increasing reduction. This may be related to the relative importance of
the noisy low lying modes in the quark propagators. 

In \cite{Glozman:2007ek} the parity-chiral group  and the effect of symmetry
breaking on the meson spectrum is discussed. E.g., whereas the $\U{1}_A$ breaking
lifts the degeneracy between pion and $a_0$ (and between $\eta$ and $f_0$) the
breaking of the chiral  $\SU{2}_L\times\SU{2}_R$ symmetry is related to the mass
differences of pion and $f_0$ (and $a_0$ and $\eta$).  From Fig.
\ref{fig:masses} we find drastic sensitivity on low modes for both, the pion 
interpolator masses and the $a_0$-mass. At low truncation levels the $a_0$-mass
rapidly  drops; it does not drop down to the pion mass value. This might indicate some
remnant of the anomaly breaking for the $J=0$ states.

The pion interpolators exhibit a puzzling behavior. The classical pion
interpolator  $\ubar\gamma_5 d$ quickly loses its exponential behavior at
larger (Euclidean) distances; only a more massive decay signal is observed at
smaller distances (\fig{RD_corr}). From truncation level 16 onwards we 
therefore do not exhibit mass values in Fig.
\ref{fig:masses} for that interpolator. A fit to the very small
time slices gives a mass approaching the
mass value from  the second interpolator $\ubar\gamma_4 \gamma_5 d$ with the
pion quantum numbers,  which couples due to PCAC (proportional to the quark
mass).

For the $J^{PC}=1^{--}$ vector meson $\rho$ there are two chiral
representations,  which correspond to the vector interpolator $\ubar \gamma_i d$
and (Dirac-)tensor interpolator $\ubar \gamma_4 \gamma_i d$
Their chiral partners
\cite{Glozman:2007ek} are the $a_1$ and the $h_1$ mesons, respectively. We did
not determine the $h_1$ mass,  since its interpolator includes disconnected graphs
(it is an $I=0$ state). There is no noticeable splitting between
the two $\rho$-interpolators for all stages of truncation. We do find,  however
intriguing behavior comparing the $\rho$-mass with the $a_1$ result. Starting
out quite differently for the full quark propagator the masses approach each
other and  are  compatible with each other from truncation level 8 onwards. 
This indicates restoration of the $\SU{2}_L\times\SU{2}_R$
symmetry for $J=1$ states. The very fact that all three interpolators
(vector,tensor and axial vector) give the same mass
hints to the restoration of the $\SU{2}_L\times\SU{2}_R\times\U{1}_A$
symmetry for $J=1$ states. The latter could be reliably
concluded, however, only after studying of the $h_1$ meson.

\section{Conclusions}

The low lying eigenvalues of the Dirac operator are usually associated with 
chiral symmetry breaking. We have computed hadron propagators while removing
increasingly more of the low lying eigenmodes of the Dirac operator. This allows
us to study their influence on certain hadron masses. Due to the relationship of
the low eigensector with chiral symmetry breaking, this amounts to partially
restoring chiral symmetry (in the valence quarks). 

We find drastic behavior for some meson interpolators when starting to remove
low  eigenmodes. At truncation level 16 the behavior saturates and then the mass
values  rise uniformly with roughly parallel slopes. The confinement properties
remain intact, i.e., we still observe clear bound states for most of the studied
isovector (scalar, axial vector and vector) mesons. An exception is the pion,
where no clear exponential decay of the correlation function is seen in the
$\ubar \gamma_5 d$ interpolator, but a massive state is seen in the $\ubar
\gamma_4 \gamma_5 d$ interpolator. The mass values of the vector meson chiral
partners $a_1$ and $\rho$ approach each other rapidly when 8 or more low
modes are removed. 

We conclude that essential confinement properties remain intact, even when the
low eigenmodes of the Dirac operator are removed in the valence sector. 
Restoration of chiral symmetry is observed in that approximation.

\begin{acknowledgments}
We would like to thank G. Colangelo and  T. DeGrand for valuable discussions.
Special thanks go to L. Glozman for helpful clarifications.  The
calculations have been performed on the SGI Altix 4700 of the
LRZ Munich and on clusters at ZID at the University of Graz. Support by
DFG SFB-TR55 and by Austrian Science Fund (FWF) DK W1203-N16
is gratefully acknowledged. M.S. is supported by the Research Executive 
Agency (REA) of the European Union under Grant Agreement 
PITN-GA-2009-238353 (ITN STRONGnet).

\end{acknowledgments}

\end{document}